\documentclass[11pt]{article}
\usepackage{lipsum}
\usepackage{neurodata}

\newcommand{\eqe}{\end{equation}}

\definecolor{MyPlum}{rgb}{0.3,0,0.3}
\definecolor{MyOrange}{rgb}{1,0.5,0}
\definecolor{deep_blue}{rgb}{0,.2,.5}
\definecolor{dark_blue}{rgb}{0,.15,.5}

\RequirePackage{amsthm}
\newtheorem{thm}{Theorem}

\newtheorem{defi}{Definition}

\newcommand{\bla}{\begin{block}}
\newcommand{\blb}{\end{block}}

\newcommand{\defa}{\begin{defi}}
\newcommand{\defb}{\end{defi}}

\newcommand{\thma}{\begin{thm}}
\newcommand{\thmb}{\end{thm}}

\newcommand{\mata}{\begin{bmatrix}}
\newcommand{\matb}{\end{bmatrix}}

\floatname{algorithm}{Procedure}

\floatname{algorithm}{Pseudocode}

\usepackage{xspace}
\usepackage{url}
\usepackage{pgfplotstable,tabularx}
\usepackage{color}
\usepackage{multicol}
\usepackage{booktabs,colortbl, array}
\usepackage{pifont}
\usepackage{pythonhighlight}

\newcommand{\graspy}{\texttt{GraSPy}\xspace}
\newcommand{\sklearn}{\texttt{scikit-learn}\xspace}
\newcommand{\networkx}{\texttt{NetworkX}\xspace}
\newcommand{\graphtool}{\texttt{graph-tool}\xspace}
\newcommand{\snap}{\texttt{Snap.py}\xspace}

\newcommand{\colwidth}{0.05\textwidth}
\newcommand{\colangle}{72}

\definecolor{MSBlue}{rgb}{.204,.353,.541}
\definecolor{slate}{rgb}{0.156,0.184,0.180}
\definecolor{lgray}{rgb}{0.312,0.368,0.360}
\definecolor{vlgray}{rgb}{0.95, 0.95, 0.95}
\definecolor{darktangerine}{rgb}{1.0, 0.66, 0.07}
\definecolor{aogr}{rgb}{0.0, 0.5, 0.0}

\newcommand{\cmark}{\ding{51}}%
\newcommand{\xmark}{\ding{55}}%

\newcommand{\greencheck}{\textcolor{aogr}\cmark}
\newcommand{\ocheck}{\textcolor{darktangerine}\cmark}

\newcommand{\redx}{\textcolor{red}\xmark}

\definecolor{tableheadcolor}{gray}{0.92}
\newcommand{\topline}{ %
        \arrayrulecolor{MSBlue}\specialrule{0.1em}{\abovetopsep}{0pt}%
        \arrayrulecolor{tableheadcolor}\specialrule{\belowrulesep}{0pt}{0pt}%
        \arrayrulecolor{MSBlue}}
\newcommand{\midtopline}{ %
        \arrayrulecolor{tableheadcolor}\specialrule{\aboverulesep}{0pt}{0pt}%
        \arrayrulecolor{MSBlue}\specialrule{\lightrulewidth}{0pt}{0pt}%
        \arrayrulecolor{white}\specialrule{\belowrulesep}{0pt}{0pt}%
        \arrayrulecolor{MSBlue}}
\newcommand{\bottomline}{ %
        \arrayrulecolor{white}\specialrule{\aboverulesep}{0pt}{0pt}%
        \arrayrulecolor{MSBlue} %
        \specialrule{\heavyrulewidth}{0pt}{\belowbottomsep}}%

\pgfplotstableset{normal/.style ={%
        header=true,
        string type,
        font=\addfontfeature{Numbers={Monospaced}}\small,
    	column type={p{.15\textwidth}},
        every odd row/.style={
            before row=
        },
        every head row/.style={
            before row={\topline\rowcolor{tableheadcolor}},
            after row={\midtopline}
        },
        every last row/.style={
            after row=\bottomline
        },
        col sep=&,
        row sep=\midrule
    }
}

\begin{document}

\title{GraSPy: Graph Statistics in Python}

\author[1, $\dagger$]{Jaewon Chung}
\author[1, $\dagger$]{Benjamin~D.~Pedigo}
\author[2]{Eric~W.~Bridgeford}
\author[1]{Bijan~K.Varjavand}
\author[3]{Hayden S. Helm}
\author[1, 3, 4, $\star$]{Joshua~T.~Vogelstein}

\affil[1]{Department of Biomedical Engineering, Johns Hopkins University, Baltimore, MD 21218}
\affil[2]{Department of Biostatistics, Johns Hopkins University of Public Health, Baltimore, MD 21218}
\affil[3]{Center for Imaging Science, Johns Hopkins University, Baltimore, MD 21218}
\affil[4]{Kavli Neuroscience Discovery Institute, Institute for Computational Medicine, Johns Hopkins University, Baltimore, MD 21218}
\affil[$\dagger$]{Denotes equal contribution}
\affil[$\star$]{Corresponding author}

\maketitle
\noindent
\textbf{
We introduce \graspy, a Python library devoted to statistical inference, machine learning, and visualization of random graphs and graph populations. This package  provides flexible and easy-to-use algorithms for analyzing and understanding graphs with a \sklearn compliant API. \graspy ~can be downloaded from Python Package Index (PyPi), and is released under the Apache 2.0 open-source license. The documentation and all releases are available at \url{https://neurodata.io/graspy}.}

\section{Introduction}
Graphs, or networks, are a mathematical representation of data that consists of discrete objects (nodes or vertices) and relationships between these objects (edges). For example, in a brain,  regions of interest can be vertices, the edges represent the presence of a structural connection between them~\citep{vogelstein2019connectal}. 
Since graphs necessarily deal with relationships between nodes,  classical statistical assumptions about independence are violated. Thus, novel  methodology is required for performing  statistical inference on graphs and populations of graphs \citep{survey-rdpg}. While the theory for inference on graphs is highly developed, to date, there has not existed a numerical package implementing these methods. \graspy fills this gap by providing implementations of algorithms with strong statistical guarantees, such as graph and multi-graph embedding methods, two-graph hypothesis testing, and clustering of vertices of graphs. Many of the algorithms implemented in \graspy are flexible and can operate on graphs that are weighted or unweighted, as well as directed or undirected. Table \ref{tab:packages} provides a comparison of \graspy to other existing graph analysis packages \citep{hagberg2008exploring, peixoto_graph-tool_2014, leskovec2016snap}.

\begin{table}[]
    \centering
    \pgfplotstabletypeset[
        header=true,
        string type,
        columns/pac/.style={column name={package}, column type={p{.19\textwidth}}},
        columns/trav/.style={column name={\rotatebox{\colangle}{traversal}}, column type={p{\colwidth}}},
        columns/net/.style={column name={\rotatebox{\colangle}{network stats.}}, column type={p{\colwidth}}},
        columns/com/.style={column name={\rotatebox{
        \colangle}{communities}}, column type={p{\colwidth}}},
        columns/embed/.style={column name={\rotatebox{\colangle}{embed}}, column type={|p{\colwidth}}},
        columns/membed/.style={column name={\rotatebox{\colangle}{mult. embed}}, column type={p{\colwidth}}},
        columns/model/.style={column name={\rotatebox{\colangle}{model fit}}, column type={p{\colwidth}}},
        columns/sim/.style={column name={\rotatebox{\colangle}{simulations}}, column type={p{\colwidth}}},
        columns/hyp/.style={column name={\rotatebox{\colangle}{hyp. test}}, column type={p{\colwidth}}},
        columns/pip/.style={column name={\rotatebox{\colangle}{pip install}}, column type={p{\colwidth}}},
        columns/plot/.style={column name={\rotatebox{\colangle}{plotting}}, column type={|p{\colwidth}}},
        every head row/.style={
            before row={
                \topline
                \rowcolor{tableheadcolor}
                & \multicolumn{3}{c}{Graph Theory} & \multicolumn{5}{c}{Statistical Modeling} & \multicolumn{2}{c}{Other}\\
                \rowcolor{tableheadcolor}
            },
            after row={\midtopline}
        },
        every odd row/.style={
            before row={\rowcolor{vlgray}}
        },
        every last row/.style={
            after row=\bottomline
        },
        col sep=&,
        row sep=\\
    ]{ pac & trav & net & com & embed & membed & model & sim & hyp  & plot & pip  \\
    \graspy 0.1.1   & \redx         & \redx & \greencheck & \greencheck & \greencheck & \greencheck & \greencheck & \greencheck & \greencheck & \greencheck  \\
    \networkx  2.3  & \greencheck   & \greencheck & \greencheck & \ocheck & \redx & \redx & \greencheck & \redx & \greencheck & \greencheck \\
    \graphtool 2.29 & \greencheck   & \greencheck & \greencheck & \ocheck & \redx & \ocheck & \greencheck & \redx & \greencheck & \redx\\
    \snap 4.1       & \greencheck   & \greencheck & \greencheck & \ocheck & \redx & \redx & \greencheck & \redx & \greencheck  & \redx \\
    }
    \caption{Qualitative comparison of Python graph analysis packages. \graspy is largely complementary to existing graph analysis packages in Python. \graspy does not implement many of the essential algorithms for operating on graphs (rather, it leverages \networkx for these implementations). The focus of \graspy is on statistical modeling of populations of networks, with features such as multiple graph embeddings, model fitting, and hypothesis testing. A  \greencheck~is given for packages that implement the respective feature, a \ocheck~for packages that partially implement the respective feature, and a \redx~is given for packages that do not implement the respective feature. Note that while a \greencheck~shows that the feature exists in the corresponding package, it does not imply that the specific algorithms are the same for all packages.}
    \vspace{-10pt}
\label{tab:packages}
\end{table}

\section{Library Overview}\label{overview}
\begin{figure}[]
    \centering
    \includegraphics[width=0.9\linewidth]{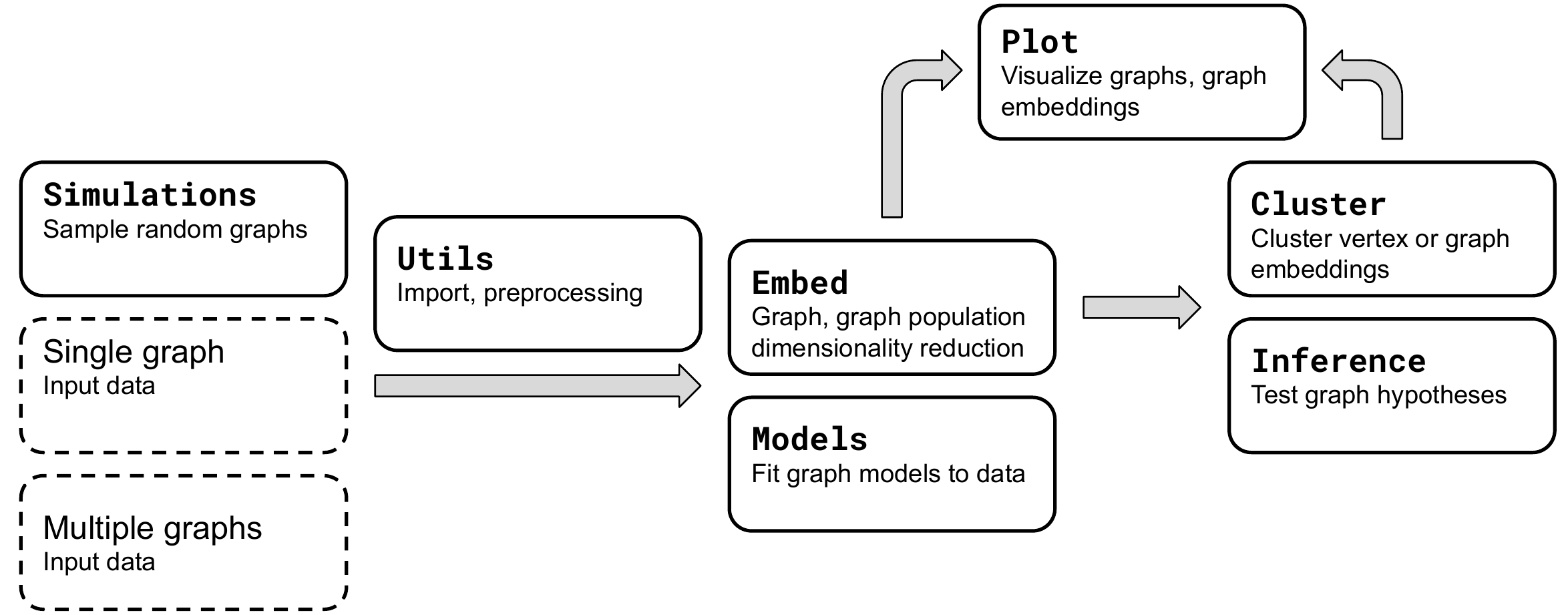}
    \caption{Illustration of modules and procedure for statistical inference on graphs, populations of graphs, or simulated data. A detailed description of each module is given in Section \ref{overview}.}
    \label{fig:graspy}
    \vspace{-10pt}
\end{figure}

\graspy includes functionality for fitting and sampling from random graph models, performing dimensionality reduction on graphs or populations of graphs (embedding), testing hypotheses on graphs, and plotting of graphs and embeddings. The following provides brief overview of different modules of \graspy. An example workflow using these modules is shown in Figure \ref{fig:graspy}. More detailed overview and code usage can be found in the tutorial section of \graspy documentation at \url{https://graspy.neurodata.io/tutorial}. All descriptions here correspond to \graspy version 0.1.1.

\begin{description}

\item \textbf{Simulations}
Several random graph models are implemented in \graspy, including the Erd\H os-R\'enyi (ER) model, stochastic block model (SBM), degree-corrected Erd\H os-R\'enyi (DCER) model, degree-corrected stochastic block model (DCSBM), and random dot product graph (RDPG) \citep{holland1983stochastic, karrer2011stochastic, young2007random}. The simulations module allows the user to sample random graphs given the parameters of one of these models. Additionally, the user can specify a distribution on the weights of graph edges.

\item \textbf{Utils}
\graspy includes a variety of utility functions for graph and graph population importing and preprocessing. \graspy can import graphs represented as \networkx objects or NumPy arrays. Preprocessing examples include finding the largest connected component of a graph, finding the intersection or union of connected components across multiple graphs, or checking whether a graph is directed.

\item \textbf{Embed}
Inferences on random graphs can leverage  low-dimensional Euclidean representations of the vertices, known as \textit{latent positions}. Adjacency spectral embedding (ASE) and Laplacian spectral embedding (LSE) are methods for embedding a single graph \citep{survey-rdpg}. Omnibus embedding and multiple adjacency spectral embedding (MASE) allows for embedding multiple graphs into the same subspace such that the embeddings can be meaningfully compared \citep{levin2017central, arroyo2019inference}. \graspy includes a method for choosing the number of embedding dimensions automatically \citep{zhu2006automatic}.

\item \textbf{Models}
\graspy includes classes for fitting random graph models to an input graph (Figure \ref{fig:models}). Currently, ER, SBM, DCER, DCSBM, and RDPG are supported for model estimation. After fitting a model to data, the model class can also output goodness-of-fit metrics (mean squared error, likelihood) and the number of estimated model parameters, allowing for model selection. The model classes can also be used to sample new simulated graphs based on the fit model. 

\begin{figure}[p]
    \centering
    \includegraphics[width=1.0\textwidth]{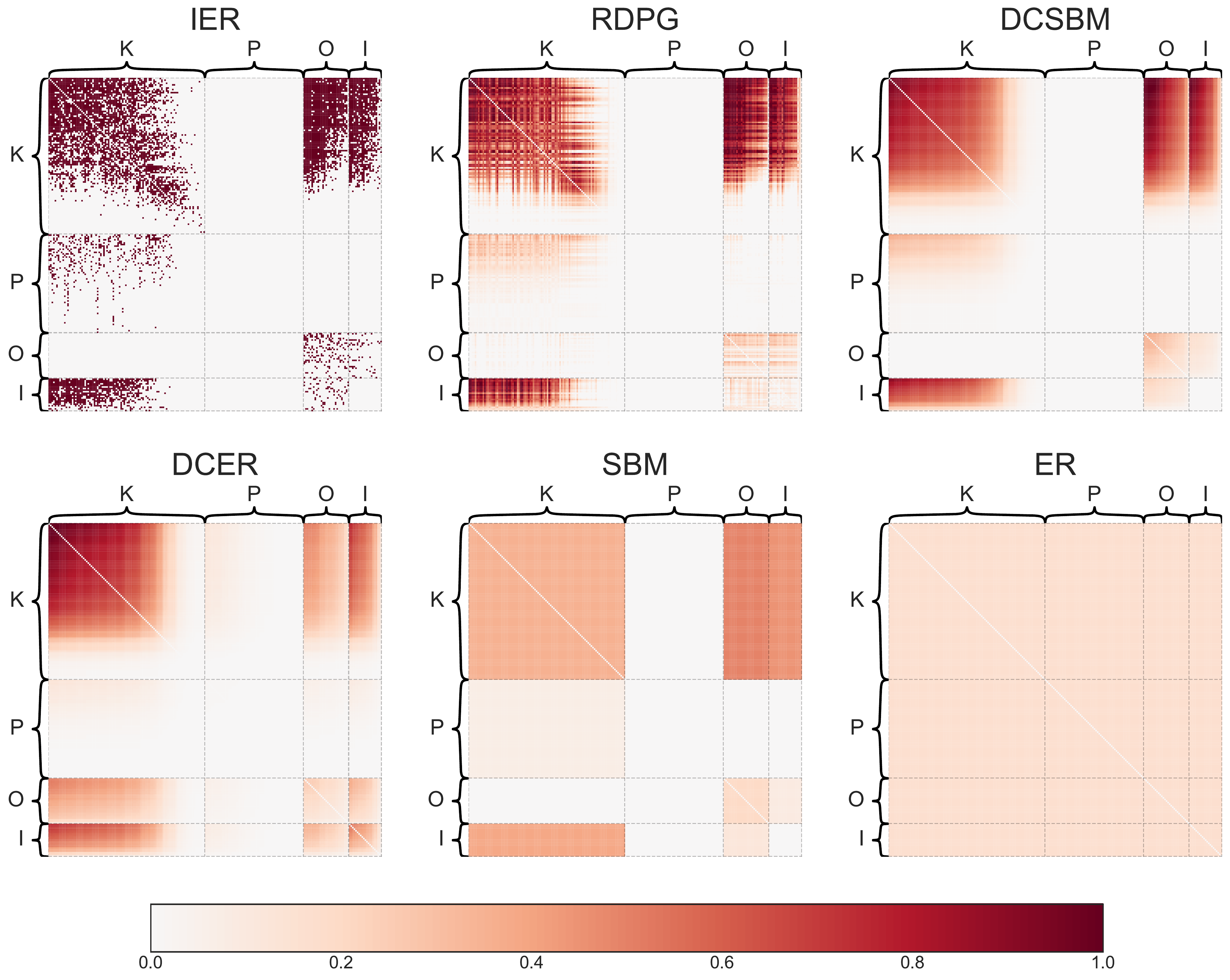}
    \caption{Connectome model fitting using \graspy. Heatmaps show the probability of potential edges for models of graphs fit to the \textit{Drosophila} larva right mushroom body connectome (unweighted, directed) \citep{eichler2017complete}. The node labels correspond to cell types: P) projection neurons, O) mushroom body output neurons, I) mushroom body input neurons. The graph models are: inhomogeneous Erd\H os-R\'enyi (IER) model in which all potential edges are specified, random dot product graph (RDPG), degree-corrected stochastic block model (DCSBM), degree-corrected Erd\H os-R\'enyi (DCER), stochastic block model (SBM), and Erd\H os-R\'enyi (ER). Blocks (cell types) are sorted by number of member vertices and nodes are sorted by degree within blocks. The code used to generate the figure is tutorial section at \url{https://neurodata.io/graspy.}
    }
    \label{fig:models}
    \vspace{-10pt}
\end{figure}

\item \textbf{Inference}
Given two graphs, a natural question to ask is whether these graphs are both random samples from the same generative distribution. \graspy provides two types of test for this null hypothesis: a latent position test and a latent distribution test. Both tests are framed under the RDPG model, where the generative distribution for the graph can be modeled as a set of latent positions. The latent position test can only be performed on two graphs of the same size and with known correspondence between the vertices of the two graphs \citep{tang2017semiparametric}. The latent distribution test can be performed on graphs without vertex alignment, or even with slightly different numbers of vertices \citep{tang2014nonparametric}.

\item \textbf{Cluster}
\graspy extends Gaussian mixture models (GMM) and k-means from \sklearn to sweep over a specified range of parameters and choose the best clustering \citep{scikit-learn}. The number of clusters and covariance structure for each GMM is chosen by Bayesian information criterion (BIC), which is a penalized likelihood function to evaluate the quality of estimators \citep{schwarz1978estimating}. Silhouette score is used to choose the number of clusters for k-means \citep{rousseeuw1987silhouettes}.  Clustering is often useful for computing the the community structure of vertices after embedding. 

\item \textbf{Plot}
\graspy extends \texttt{seaborn} to visualize graphs as adjacency matrices and embedded graphs as paired scatter plots \citep{seaborn}. Individual graphs can be visualized using heatmap function, and multiple graphs can be overlaid on top of each other using gridplot. The nodes in both graph visualizations can be sorted by various node metadata, such as node degree or assigned node labels. Pairplot can visualize high dimensional data, such as embeddings, as a pairwise scatter plot.

\end{description}

\newpage
\section{Code example}
Given the connectomes of the \textit{Drosophila} larva left and right mushroom bodies, one natural question to ask is: how similar are these graphs \citep{eichler2017complete}? We can frame this question as whether these graphs are generated from the same distribution of latent positions \citep{tang2014nonparametric}. We can use the latent distribution test to test this hypothesis: 



\begin{python}[caption={Python}]
from graspy.datasets import load_drosophila_left, load_drosophila_right
from graspy.inference import LatentDistributionTest

# Load data
left_graph = load_drosophila_left()
right_graph = load_drosophila_right()

# Initialize hypothesis test object and compute p-value
ldt = LatentDistributionTest(n_components=3, n_bootstraps=500)
p_value = ldt.fit(left_graph, right_graph)
print("p-value: " + str(p_value))
p-value: 0.002
\end{python}

\section{Conclusion}
\graspy is an open-source Python package to perform  statistical analysis on graphs and graph populations. Its compliance with the \sklearn API makes it an easy-to-use tool for anyone familiar with machine learning in Python \citep{sklearn_api}. In addition, \graspy is implemented with an extensible class structure, making it easy to modify and add new algorithms to the package. As \graspy continues to grow and add functionality, we believe it will accelerate statistically principled discovery in any field of study concerned with graphs or populations of graphs. 

\section*{Acknowledgements}
This work is graciously supported by the DARPA, under agreement numbers FA8650-18-2-7834 and FA8750-17-2-0112. We thank all the contributors for assisting with writing \graspy. We thank the NeuroData Design class, the NeuroData lab, and Carey E.~Priebe at Johns Hopkins University for helpful feedback.




\nocite{*}

\pagebreak
\vspace{5mm}
\bibliography{neurodata}
\bibliographystyle{IEEEtran}

\end{document}